\documentclass[twocolumn,showpacs,floats,floatfix,superscriptaddress,aps,pra]{revtex4}

\usepackage{mathptmx}
\usepackage{subfigure}
\usepackage{psfrag,graphicx}
\usepackage{dcolumn}
\usepackage{amsmath,amssymb}
\usepackage{bm}
\usepackage{color}
\usepackage{latexsym}
\usepackage{epstopdf}
\usepackage{color}
\usepackage[english]{babel}
\usepackage{latexsym}
\usepackage{psfrag,graphicx}
\usepackage{subfigure}
\usepackage{amsmath}
\usepackage{amssymb}
\usepackage{amsfonts}
\usepackage{bm}
\usepackage{natbib}
\usepackage{epstopdf}
\usepackage{appendix}

\definecolor{mygrey}{gray}{0.35}
\definecolor{myblue}{rgb}{0.2,0.2,0.8}
\definecolor{myzard}{cmyk}{0,0,0.05,0}
\definecolor{mywhite}{rgb}{1,1,1}
\definecolor{mywhite}{rgb}{1,1,1}
\definecolor{myred}{rgb}{1,0.,0.3}

\usepackage[colorlinks=true,citecolor=myblue,linkcolor=myred]{hyperref}

\def\ba{\begin{align}}
\def\enda{\end{align}}
\def\bi{\begin{itemize}}
\def\ei{\end{itemize}}

\def\be{\begin{equation}}
\def\ee{\end{equation}}
\def\bea{\begin{eqnarray}}
\def\eea{\end{eqnarray}}
\def\bse{\begin{subequations}}
\def\ese{\end{subequations}}

\begin{document}

\author{Peter A. Ivanov}
\affiliation{Department of Physics, Sofia University, James Bourchier 5 blvd, 1164 Sofia, Bulgaria}
\author{Naoum I. Karchev}
\affiliation{Department of Physics, Sofia University, James Bourchier 5 blvd, 1164 Sofia, Bulgaria}
\author{Nikolay V. Vitanov}
\affiliation{Department of Physics, Sofia University, James Bourchier 5 blvd, 1164 Sofia, Bulgaria}
\author{Dimitris G. Angelakis}
\affiliation{School of Electronic and Computer Engineering, Technical University of Crete, Chania, Crete, 73100 Greece}
\affiliation{Center for Quantum Technologies, National University of Singapore, 2 Science Drive 3, Singapore 117543}
\title{Quantum simulation of superexchange magnetism in linear ion crystals }

\begin{abstract}
We present a system for the simulation of Heisenberg models with spins $s=\frac{1}{2}$ and $s=1$ with a linear crystal of trapped ions.
We show that the laser-ion interaction induces a Jaynes-Cummings-Hubbard interaction between the atomic V-type level structure and the two phonon species.
In the strong-coupling regime the collective atom and phonon excitations become localized at each lattice site and form an effective spin system with varying length.
We show that the quantum-mechanical superexchange interaction caused by the second-order phonon hopping processes creates a Heisenberg-type coupling between the individual spins.
Trapped ions allow to control the superexchange interactions by adjusting the trapping frequencies, the laser intensity, and the detuning.
\end{abstract}

\maketitle

\section{Introduction}
The current ion trap technology is among the most promising physical systems for the implementation of quantum simulator of many-body models \cite{Cirac_2012,Johanning_2009}, such as quantum spin magnetism \cite{Friedenauer_2008,Kim_2010,Cohen} and quantum structural phase transitions \cite{Porras_2012,Bermudez_2012}, by means of spin-dependent force.
Recent experimental realization of the quantum phase transition from localized Mott insulator state to a delocalized superfluid state of polaritonic excitations in a system of trapped ions opened fascinating prospects to explore strongly correlated spin-boson systems under controlled conditions \cite{Toyoda2013}.
Such a quantum phase transition of hybrid light matter excitations is described by the celebrated Jaynes-Cummings-Hubbard model (JCH) in which the elementary excitations are polariton quasiparticles.
Originally the JCH model was introduced to describe a coupled array of electromagnetic resonators, each coupled coherently to a two-level system \cite{Angelakis_pra2007,Greentree_np}.
The strong coupling between the bosonic mode and the atomic levels introduces a nonlinearity into the system, leading to an effective repulsive photon-photon interaction \cite{Hartmann}.
With trapped ions the bosonic mode is naturally provided by the quantized radial ion oscillations, which we refer to as local phonons, while the coupling between the phonons and the two-state atom is provided by an external laser field \cite{Ivanov2009}.

\begin{figure}[tb]
\includegraphics[width=0.35\textwidth]{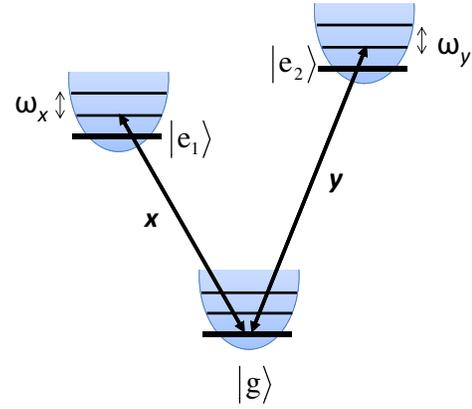}
\caption{The V-type three-level system consists of the ground state $|g\rangle$ and two metastable excited states $|e_{1}\rangle$, $|e_{2}\rangle$.
Two laser beams with properly chosen frequencies and polarizations create the JC couplings between the V-type level structure and the two radial $x$ and $y$ phonon species.}
\label{fig1}
\end{figure}

In the strongly coupled regime, the on-site repulsion dominates over the hopping processes and hence the system is in the Mott phase, in which the polaritonic excitations become localized in each lattice site \cite{Tomadin,Hartmann_LS,Mering2009}.
It was shown that in this phase a useful mapping to spin-$\frac{1}{2}$ XX model is possible by considering the low-lying energy states of two Mott-insulating lobes \cite{Angelakis_pra2007}.
Because the Mott state of the JCH model is not degenerate, the perturbative hopping processes in the single Mott-insulating lobe do not introduce additional spin dynamics.
The situation is changed significantly when the three-level atom is coupled to two bosonic species via Jaynes-Cummings (JC) interaction, see Fig. \ref{fig1}.
In this case for unit filling factor the ground-state is double degenerate such that the low-energy physics is described by effective spin-$\frac{1}{2}$ system. Moreover, due to the second-order hopping processes the effective spins on different lattice sites become coupled by Heisenberg exchange interaction \cite{Ji_prl2007,Angelakis_epr2008}.

In this paper, we propose a quantum simulation of anisotropic Heisenberg spin models with spins $s=\frac{1}{2}$ and $s=1$ in a system of trapped ions.
The underlying idea is based on the mapping of the Jaynes-Cummings-Hubbard model in a V-type three-level system (JCHv) to an effective Heisenberg spin model using a linear ion crystal.
The two-bosonic species in the JCHv model are represented by the two radial local phonon modes, while the long-range phonon hopping dynamics appear naturally due to the Coulomb interaction. We shall show that the laser beams in two orthogonal directions tuned near the respective red sideband transition can be used to provide the JC couplings between the three internal states of the ion and the two radial phonon species.
Another possible realization of the JCHv model is based on an oscillating magnetic field gradient, where the JC couplings are controlled by the magnetic gradient.
When the phonon hopping dynamics is suppressed the second-order virtual processes can induce an effective Heisenberg exchange between the localized polaritonic excitations in different lattice sites.
The nature of the conserved polariton quasiparticles can be transformed into atomic or phononic excitations by controlling the laser intensity and detuning.
We will show that in the strongly coupled regime the Heisenberg spin models with $s=\frac{1}{2}$ and $s=1$ can be realized.
As for the ultracold two-component atoms in an optical lattice \cite{Trotzky_2008,Kuklov_2003,Duan_2003,Altman_2003}, we show that a higher-order virtual phonon hopping processes in both radial directions mediate the spin-spin interactions.
We calculate the respective tunneling matrix elements in the case of anisotropic spin-phonon couplings and detuned JC interaction. We consider two cases.
(i) In the case of one excitation per lattice site the corresponding spin dynamics is described by the anisotropic XXZ Heisenberg model in the presence of external effective magnetic field.
We show that the anisotropy in the system can be controlled by the external parameters such as the laser intensity and the detuning, which allows us to realize an easy-axis or easy-plane ferromagnet.
(ii) For the two excitations per lattice site, the underlying lowest energy physics of the JCHv model is described by an effective spin $s=1$ system.
We show that the spin-spin interaction induced by the second-order hopping events is governed by the highly anisotropic spin $s=1$ Heisenberg model.
Such a spin-1 model can serve as a test bed to explore a novel topological orders.

The paper is organized as follows.
For the sake of reader's convenience, in Sec. \ref{PH} we introduce the tight-binding model which describes the dynamics of the local radial phonons in the linear ion crystal.
In Sec. \ref{Implementation} we provide a scheme for the realization of the JCHv model with a laser driven linear ion crystal.
In Sec. \ref{EnS} we discuss the relevant energy scales of the JCHv Hamiltonian and the perturbative approach that incorporates the effect of the phonon hopping.
In Sec. \ref{Spin1/2} we discuss the realization of the XXZ spin $s=\frac{1}{2}$ Heisenberg model in the case of anisotropic spin-phonon couplings.
In Sec. \ref{Spin1} we derive the effective spin $s=1$ Heisenberg-like Hamiltonian. Finally, the conclusions is presented in Sec. \ref{conclusion}.

\section{Phonon Hamiltonian}\label{PH}

\begin{figure}[tb]
\includegraphics[width=0.45\textwidth]{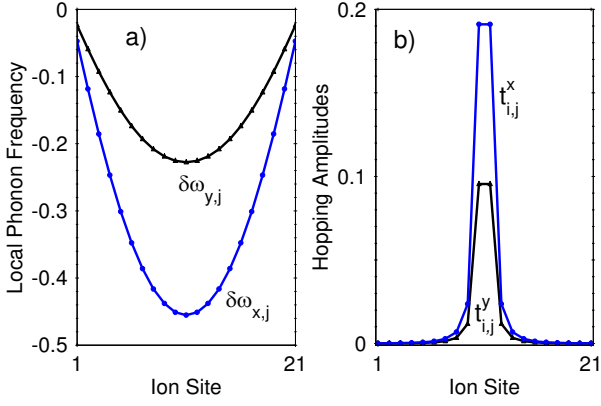}
\caption{(Color online) a) The on-site phonon frequencies $\delta\omega_{\beta,j}$ and b) Coulomb-mediated long-range hopping amplitudes $t_{11,j}^{\beta}$ in the radial $x$ and $y$ directions in units of $\omega_{z}$ for a linear ion crystal with $N=21$ ions as a function of the ion positions. The aspect ratios are $\omega_{x}/\omega_{z}=50$ and $\omega_{y}/\omega_{z}=100$.}
\label{fig0}
\end{figure}

We consider a crystal of $N$ identical ions with charge $e$ and mass $m$ confined in a Paul trap along the $z$ axis with trap frequencies $\omega_{\alpha}$ ($\alpha=x,y,z$).
The potential energy $\hat{V}$ of the trapped ions is a sum of the effective harmonic potential and the mutual Coulomb repulsion between the ions of the trap,
\begin{equation}
\hat{V}=\frac{m}{2}\sum_{\alpha}\sum_{j=1}^{N}\omega_{\alpha}\hat{r}_{\alpha,j}^{2}+\sum_{j>k}^{N}\frac{e^{2}}{|\hat{\vec{r}}_{j}-
\hat{\vec{r}}_{k}|},\label{V}
\end{equation}
where $\hat{\vec{r}}_{j}$ is the position vector operator of ion $j$.
For sufficiently low temperature and strong radial confinement ($\omega_{x,y}\gg \omega_{z}$) the ions are arranged in a linear configuration and occupy equilibrium positions $z_{i}^{0}$ along the trapping $z$ axis, which are determined by the minimization of potential (\ref{V}) \cite{James}. The position operator of ion $j$ is
\begin{equation}
\hat{\vec{r}}_{j}=(z_{j}^{0}+\delta \hat{r}_{z,j})\vec{e}_{z}+\delta \hat{r}_{x,j}\vec{e}_{x}+\delta \hat{r}_{y,j}\vec{e}_{y},
\end{equation}
where $\delta \hat{r}_{\alpha,j}$ are the displacement operators around the equilibrium positions. We expand the potential $\hat{V}$ for small displacement around the equilibrium positions and consider the motion only in the radial $x$-$y$ plane which gives ($\beta=x,y$ and $\hbar=1$ from now on)
\begin{eqnarray}
\hat{H}_{xy}&=&\frac{1}{2m}\sum_{j}\sum_{\beta}\hat{p}_{\beta,j}^{2}+\frac{m}{2}\sum_{j}\sum_{\beta}
\omega_{\beta}^{2}\delta \hat{r}_{\beta,j}^{2}\notag\\
&&-\sum_{\beta}\sum_{j>k}\frac{e^{2}}{2|z_{j}^{0}-z_{k}^{0}|^{3}}(\delta \hat{r}_{\beta,k}-\delta \hat{r}_{\beta,j})^{2}.
\end{eqnarray}
In the following we treat each ion as an individual oscillator by introducing creation $\hat{a}_{\beta,j}^{\dag}$ and annihilation $\hat{a}_{\beta,j}$ operators of local phonons at site $j$ and direction $\beta$, such that $\hat{p}_{\beta,j}=i\sqrt{m\omega_{\beta}/2}(\hat{a}_{\beta,j}^{\dag}-\hat{a}_{\beta,j})$ and $\delta \hat{r}_{\beta,j}=(\hat{a}_{\beta,j}^{\dag}+\hat{a}_{\beta,j})/\sqrt{2m\omega_{\beta}}$, respectively.
Assuming that the radial trapping potential is much larger than the Coulomb interaction we arrive at \cite{Porras2004,Ivanov2009}
\begin{eqnarray}
&&\hat{H}_{xy}=\hat{H}_{0}+\hat{H}_{\rm b},\notag\\
&&\hat{H}_{0}=\sum_{\beta}\sum_{j}\omega_{\beta,j}\hat{a}_{\beta,j}^{\dag}\hat{a}_{\beta,j},\notag\\
&&\hat{H}_{\rm b}=\sum_{j>k}t_{j,k}^{\beta}(\hat{a}_{\beta,j}^{\dag}\hat{a}_{\beta,k}
+\hat{a}_{\beta,j}\hat{a}_{\beta,k}^{\dag}),\label{Hxy}
\end{eqnarray}
where the fast rotating terms $\hat{a}_{\beta,j}^{2}$ and $(\hat{a}_{\beta,j}^{\dag})^{2}$ are neglected, which is justified as long as $\omega_{\beta}\gg t_{j,k}^{\beta}$. Here $\hat{H}_{0}$ is the free bosonic term with renormalized phonon frequency $\omega_{\beta,j}=\omega_{\beta}+\delta\omega_{\beta,j}$,
\begin{equation}
\delta\omega_{\beta,j}=-\sum_{k\neq j}\frac{e^{2}}{2m\omega_{\beta}|z_{j}^{0}-z_{k}^{0}|^{3}},
\end{equation}
which is caused by the interaction of ion $j$ with the rest of the ion crystal.
The term $\hat{H}_{\rm b}$ describes the Coulomb mediated long-range phonon hopping dynamics with hopping strength
\begin{equation}
t_{j,k}^{\beta}=\frac{e^{2}}{2m\omega_{\beta}|z_{j}^{0}-z_{k}^{0}|^{3}}.
\end{equation}
The distribution of the hopping values and the local phonon frequencies are shown in Fig. \ref{fig0} for a chain of $21$ ions. We note that the phonon hopping dynamics subject to the tight-binding Hamiltonian (\ref{Hxy}) was experimentally observed recently in a linear Paul trap \cite{Haze_2012} as well as with trapped ions in a double-well potential \cite{Harlander,Brown}.

In the following section we show that the laser-ion interaction induces a coupling between the V-type internal ion states and the two radial local phonon species.

\section{Jaynes-Cummings-Hubbard model in a V-system}\label{Implementation}

Trapped ions are a suitable system to implement JC
interaction in a two-level system by driving a red-sideband transition with an external laser \cite{Wineland} or a magnetic-field gradient \cite{Mintert,Ospelkaus_prl2009}.
The two-level system typically consists of two metastable levels.
The JC interaction can be created by a direct two-photon optical transition, as in the $4s{}^2S_{1/2}-3d{}^2D_{5/2}$ transition in $^{40}$Ca$^{+}$ ion or alternatively one can use radio-frequency, or hyper-fine levels where the JC coupling is driven by Raman-type interaction.
Here we consider an atomic V-type system, which consists of a ground state $|g\rangle$ and two metastable levels $|e_{1}\rangle$ and $|e_{2}\rangle$ with transition frequencies $\omega_{e,1}$ and $\omega_{e,2}$, which are depicted in Fig. \ref{fig1}.
For example, such a level structure occurs in $^{40}$Ca$^{+}$ ion with ground state $|g\rangle=|S_{1/2},m_{J}=-1/2\rangle$ and two excited levels $|e_{1}\rangle=|D_{5/2},m_{J}=-5/2\rangle$ and $|e_{2}\rangle=|D_{5/2},m_{J}=-3/2\rangle$ \cite{Kaler_2003}.
We assume that the linear ion crystal interacts with two laser beams along the two orthogonal radial directions with laser frequencies $\omega_{{\rm L},x}$ and $\omega_{{\rm L},y}$.
The Hamiltonian describing the laser-ion interaction after making the optical rotating-wave approximation is given by \cite{Wineland}
\begin{eqnarray}
\hat{H}&=&\hat{H}_{xy}+\Omega_{x}\sum_{j}\{|e_{1,j}\rangle\langle g_{j}|e^{i\eta_{x}(\hat{a}_{x,j}^{\dag}+\hat{a}_{x,j})-i\delta_{x}t}+{\rm H.c.}\}\notag\\
&&+\Omega_{y}\sum_{j}\{|e_{2,j}\rangle\langle g_{j}|e^{i\eta_{y}(\hat{a}_{y,j}^{\dag}+\hat{a}_{y,j})-i\delta_{y}t}+{\rm H.c.}\}.\label{H_Laser}
\end{eqnarray}
Here $\Omega_{\beta}$ is the Rabi frequency and $\eta_{\beta}=|\vec{k}_{\beta}|/\sqrt{2m\omega_{\beta}}$ is the Lamb-Dicke parameters along the $\beta$ axis,
 with $\vec{k}_{\beta}$ being the laser wave vector.
$\delta_{x}=\omega_{{\rm L},x}-\omega_{e,1}$, $\delta_{y}=\omega_{{\rm L},y}-\omega_{e,2}$ are the laser detunings.
We assume that the laser frequencies are tuned near the motional red sideband along the two radial directions,
\bse
\begin{align}
\omega_{{\rm L},x} &=\omega_{e,1}-\omega_{0}-(\omega_{x}-\Delta_{x}),\\
\omega_{{\rm L},y} &=\omega_{e,2}-\omega_{0}-(\omega_{y}-\Delta_{y}),
\end{align}
\ese
where the conditions $\Delta_{\beta},\omega_{0}\ll \omega_{\beta},\omega_{e,1(2)}$ are satisfied.
The detunings $\Delta_{\beta}=\Delta-\delta\omega_{\beta}$ introduce effective trapping frequencies along the two orthogonal directions,
 while the detuning $\omega_{0}$ introduces an effective spin frequency.
The Hamiltonian (\ref{H_Laser}) after transforming into a rotating frame with respect to
\begin{align}
\hat{U}(t) = \exp \Big[i\sum_{j}\{\sum_{a=1}^{2}\omega_{0}|e_{a,j}\rangle \langle e_{a,j}|
  -\sum_{\beta}(\omega_{\beta}-\Delta_{\beta})\hat{a}_{\beta,j}^{\dag}\hat{a}_{\beta,j}\}t \Big],
\end{align}
in the Lamb-Dike limit and the vibration rotating-wave approximation, reads
\bse\label{JCH}
\begin{align}
\hat{H}_{\rm JCHv} &= \hat{H}_{\rm JC}+\hat{H}_{\rm b},\\
\hat{H}_{\rm JC} &= \sum_{j}\Big[\sum_{\beta}\Delta_{\beta,j}\hat{a}_{\beta,j}^{\dag}\hat{a}_{\beta,j}+\omega_{0}(|e_{1,j}\rangle\langle e_{1,j}|+|e_{2,j}\rangle\langle e_{2,j}|) \notag\\
& +g_{x}(\hat{a}_{x,j}|e_{1,j}\rangle \langle g_{j}|+{\rm H.c})+g_{y}(\hat{a}_{y,j}|e_{2,j}\rangle \langle g_{j}|+{\rm H.c})\Big],
\end{align}
\ese
where $\hat{H}_{\rm JCHv}=\hat{U}^{\dag}\hat{H}\hat{U}-i\hat{U}^{\dag}\partial_{t}\hat{U}$. Here $g_{\beta}=\eta_{\beta}\Omega_{\beta}$ are the spin-phonon couplings and $\Delta_{\beta,j}=\Delta+\delta\omega_{\beta,j}-\delta\omega_{\beta}$.
The term $\hat{H}_{\rm JC}$ describes the JC model in a V-type atomic system, where the first two terms correspond to the effective energies of the local phonons and ions,
 while the last two terms describe the couplings between the internal levels and the $x$ and $y$ local phonons.
The term $\hat{H}_{\rm b}$ describes the nonlocal hopping of the two-phonon species between different lattice sites and allow us direct comparison with the case of two-component ultracold atom gas in an optical lattice \cite{Trotzky_2008,Kuklov_2003,Duan_2003,Altman_2003}.
Finally we note that the continuous U(1) symmetry of the Hamiltonian (\ref{JCH}) associated with the conservation of the total number of excitations is generated by the excitation operator $\hat{N}=\sum_{j}\hat{N}_{j}$ with $\hat{N}_{j}=\sum_{\beta}\hat{a}_{\beta,j}^{\dag}\hat{a}_{\beta,j}+\sum_{a=1,2}|e_{a,j}\rangle\langle e_{a,j}|$.

Alternatively, the JCHv model can be implemented by using an oscillating magnetic field gradient.
Consider for example the V-type level structure of $^{171}$Yb$^{+}$ ion, which consists of the ground state $|g\rangle=|F=0\rangle$ and the two metastable excited levels $|e_{1}\rangle=|F=1,m_{F}=1\rangle$ and $|e_{2}\rangle=|F=1,m_{F}=-1\rangle$.
In that case an oscillating magnetic field gradient along the two orthogonal directions can be used to create the desired JC couplings.
In the Appendix \ref{Impl_LI} we provide the scheme for the realization of the JCHv model using a magnetic-field gradient.

\section{Energy Scales}\label{EnS}

\begin{figure}[tb]
\includegraphics[width=0.45\textwidth]{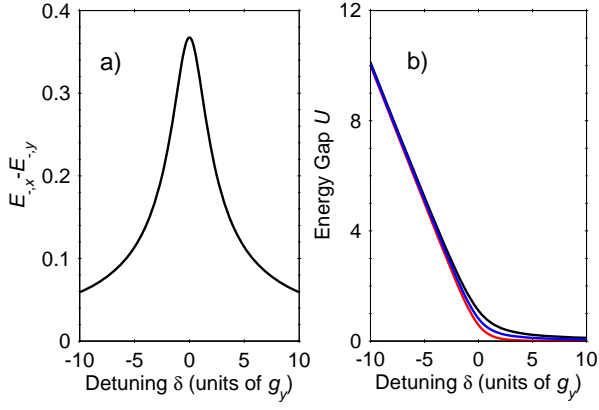}
\caption{(Color online) a) The energy splitting $E_{-,x}-E_{-,y}$ versus $\delta$. The spin-phonon couplings satisfy $g_{y}=\sqrt{2.5}g_{x}$. b) The energy difference $U=E_{\rm exc}-2E_{-,y}$, versus the detuning $\delta$. Here $E_{\rm exc}$ is the energy for the state with two excitations in one site and none in another. The three curves are the energy differences with respect to the three low-energy states.}
\label{fig2}
\end{figure}

Similar to the two-level JCH model, the three-level JCHv model is not generally amenable to an exact solution.
The particular limit, which we study in the present paper is the strong-coupling regime $g_{\beta}\gg t_{i,j}^{\beta}$, which allows us to diagonalize the term $\hat{H}_{\rm JC}$ in (\ref{JCH}) and then treat the hopping term $\hat{H}_{\rm b}$ as a perturbation.
Because the number of excitations $\hat{N}_{j}$ in each site $j$ is a constant of motion the Hilbert space is decomposed in subspaces with well-defined numbers of excitations.
In the following we consider the homogeneous limit $\Delta_{\beta,j}\approx\Delta$ ($\Delta\gg \delta\omega_{\beta,j}-\delta\omega_{\beta}$) but in the numerical simulations we take into account the finite-size effects.
For null excitations the ground state of $\hat{H}_{\rm JC}$ is nondegenerate and given by $|g,0_{x},0_{y}\rangle$ with $E_{0}=0$.
Here the state $|l,n_{x},n_{y}\rangle$ ($l=g,e_{1},e_{2}$) describes an ion in the internal state $|l\rangle$ together with $n_{x}$ and $n_{y}$ local phonons.
For one excitation per lattice site (unit filling factor) the energy spectrum is
\be
E_{\pm,\beta}=\Delta+\frac{\delta}{2}\pm\sqrt{\frac{\delta^{2}}{4}+g_{\beta}^{2}},
\ee
 with $\delta=\omega_{0}-\Delta$.
The dressed eigenstates corresponding to the two lowest eigenfrequencies are
\bse\label{ES}
\begin{align}
\left|\uparrow\right\rangle &=\left|-\right\rangle_{x}=\cos\theta_{x}|g,1_{x},0_{y}\rangle-\sin\theta_{x}|e_{1},0_{x},0_{y}\rangle,\\
\left|\downarrow\right\rangle &=\left|-\right\rangle_{y}=\cos\theta_{y}|g,0_{x},1_{y}\rangle-\sin\theta_{y}|e_{2},0_{x},0_{y}\rangle,
\end{align}
\ese
where the mixing angle is defined by
\begin{equation}
\theta_{\beta} = \tan^{-1}\frac{2g_{\beta}}{\delta+\sqrt{\delta^{2}+4g_{\beta}^{2}}}.
\end{equation}
In the strong-coupling regime the energy splitting $E_{+,\beta}-E_{-,\beta}$ is large compared to any other energy scale in the system;
hence the two low energy states (\ref{ES}) can be treated as an $s=\frac{1}{2}$ effective spin system with the energy difference shown in Fig. \ref{fig2}(a).
The two states become degenerate for $g_{x}=g_{y}$, while for unequal couplings there is a finite energy difference, which tends to zero for large detuning $\delta$.
The eigenstates (\ref{ES}) describe the polaritonic excitation in the system caused by the strong spin-phonon coupling.
The nature of the polaritonic excitations can be controlled by the external parameters, such as laser intensity and detuning.
For instance, in the limit of large negative detuning ($|\delta|\gg g_{\beta}$) the polaritons are transformed into atomic excitations, $\left|\uparrow\right\rangle\approx|e_{1},0_{x},0_{y}\rangle$, $\left|\downarrow\right\rangle\approx|e_{2},0_{x},0_{y}\rangle$, while for large positive detuning ($\delta\gg g_{\beta}$) the excitations become purely phononic, $\left|\uparrow\right\rangle\approx|g,1_{x},0_{y}\rangle$ and $\left|\downarrow\right\rangle\approx|g,0_{x},1_{y}\rangle$.

We can extend the discussion to the case of integer filling of two excitations per site. In that case the lowest energy Hilbert space of a lattice site consists of three eigenstates given by
\bse\label{n3}
\begin{align}
|1\rangle =&\cos\theta_{2,x}|g,2_{x},0_{y}\rangle-\sin\theta_{2,x}|e_{1},1_{x},0_{y}\rangle,\\
|0\rangle =&\cos\varphi|g,1_{x},1_{y}\rangle-\sin\varphi(\sin\zeta |e_{1},0_{x},1_{y}\rangle\\
&+\cos\zeta|e_{2},1_{x},0_{y}\rangle),\notag\\
\left|-1\right\rangle =&\cos\theta_{2,y}|g,0_{x},2_{y}\rangle-\sin\theta_{2,y}|e_{2},0_{x},1_{y}\rangle,
\end{align}
\ese
where the mixing angles are defined as
\bse
\begin{align}
\theta_{2,\beta} &=\tan^{-1}\frac{\sqrt{2}g_{\beta}}{\frac{\delta}{2}+\sqrt{2g_{\beta}^{2}+\frac{\delta^{2}}{4}}},\\
\varphi &=\tan^{-1}\frac{\sqrt{g_{x}^{2}+g_{y}^{2}}}{\frac{\delta}{2}+\sqrt{g_{x}^{2}+g_{y}^{2}+\frac{\delta^{2}}{4}}},\\
\zeta &=\tan^{-1}\frac{g_{x}}{g_{y}}.
\end{align}
\ese
The corresponding energies of the states (\ref{n3}) are
\bse
\begin{align}
E_{1} &=2\Delta+\frac{\delta}{2}-\sqrt{2g_{x}^{2}+\frac{\delta^{2}}{4}},\\
E_{0} &=2\Delta+\frac{\delta}{2}-\sqrt{g_{x}^{2}+g_{y}^{2}+\frac{\delta^{2}}{4}},\\
E_{-1} &=2\Delta+\frac{\delta}{2}-\sqrt{2g_{y}^{2}+\frac{\delta^{2}}{4}},
\end{align}
\ese
respectively. By using the same arguments as above, we conclude that in the strong-coupling regime the eigenstates (\ref{n3}) represent an effective spin $s=1$ system.
Again, the nature of the polaritonic excitations can be transformed into various kinds depending on the spin-phonon couplings $g_{\beta}$ and the detuning $\delta$.
For large negative detuning ($|\delta|\gg g_{\beta}$) the spin states contain one atomic excitation and one phonon excitation, $|1\rangle\approx|e_{1},1_{x},0_{y}\rangle$, $|0\rangle\approx(\sin\zeta |e_{1},0_{x},1_{y}\rangle+\cos\zeta|e_{2},1_{x},0_{y}\rangle)/\sqrt{2}$ and $\left|-1\right\rangle\approx|e_{2},0_{x},1_{y}\rangle$, while in the limit $\delta\gg g_{\beta}$ the atomic transitions are suppressed so that the spin states contain only two phononic excitations, $|1\rangle\approx|g,2_{x},0_{y}\rangle$, $|0\rangle\approx|g,1_{x},1_{y}\rangle$ and $\left|-1\right\rangle\approx|g,0_{x},2_{y}\rangle$.
In general for $n$ excitations the low energy manifold consists of $n+1$ eigenstates, which make it possible to simulate spin $\frac{1}{2}n$ particles \cite{Angelakis_epr2008}.

Now we examine the effect of the finite hopping amplitudes $t_{i,j}^{\beta}$.
First, we note that the energy spectrum of $\hat{H}_{\rm JC}$ displays a particle-hole gap, which implies that there exists an energy difference $U$ between the states with $n$ excitations per site and the states with $n+1$ excitations in one site and $n-1$ in another \cite{Mering2009}.
In Fig. \ref{fig2}(b), we plot these energy differences for $n=1$. For large positive detuning the energy gap becomes vanishingly small, while in the limit of large negative detuning the gap scales as $U\sim |\delta|$ \cite{Tomadin}.

As long as the energy gap is much higher than the hopping strength $t_{i,j}^{\beta}$ ($U\gg t_{i,j}^{\beta}$) the excitations are strongly localized in each site, so that the system is in the Mott insulator phase.
In this regime a single excitation jump changes the total on-site polaritonic excitations and therefore such processes are highly suppressed. Although the hopping events are frozen, the spin degrees of freedom can be coupled by an effective superexchange interaction.
Indeed, the next high-lying states containing $n+1$ or $n-1$ excitations can be reached as virtual intermediate states in second-order hopping processes. Such second-order hopping events mediate the spin-spin interaction between the effective spin systems on different sites and can be studied using the expression
\begin{eqnarray}
(\hat{H}_{\rm eff})_{r_{j}r_{k}^{\prime},d_{j}d_{k}^{\prime}}&=&(\hat{H}_{\rm JC})_{r_{j}r_{k}^{\prime},d_{j}d_{k}^{\prime}}+\sum_{\chi}\langle r_{j},r_{k}^{\prime}|\hat{H}_{\rm b}|\chi\rangle\langle \chi|\hat{H}_{\rm b}|d_{j},d_{k}^{\prime}\rangle\notag\\
&&\times\frac{1}{2}\left(\frac{1}{E_{r_{j}r_{k}^{\prime}}-E_{\chi}}
+\frac{1}{E_{d_{j}d_{k}^{\prime}}-E_{\chi}}\right).\label{PertH}
\end{eqnarray}
The matrix elements of the effective Hamiltonian (\ref{PertH}) describe the coupling between the spin states $|r_{j},r^{\prime}_{k}\rangle\leftrightarrow |d_{j},d^{\prime}_{k}\rangle$ on sites $j$ and $k$ with energies $E_{r_{j},r_{k}^{\prime}}$ and $E_{d_{j},d_{k}^{\prime}}$, respectively, created via hopping processes to state $|\chi\rangle$ with energy $E_{\chi}$ which contains $n+1$ excitations in one site, and $n-1$ in another.
In the following we will consider only the spin-$\frac{1}{2}$ and spin-1 models; then the spin indices take values $r,d=\uparrow, \downarrow$ for $s=\frac{1}{2}$ or $r,d=1,0,-1$ for $s=1$.

\section{Spin-$\frac{1}{2}$ Anisotropic XXZ Heisenberg model}\label{Spin1/2}

\begin{figure}[tb]
\includegraphics[width=0.45\textwidth]{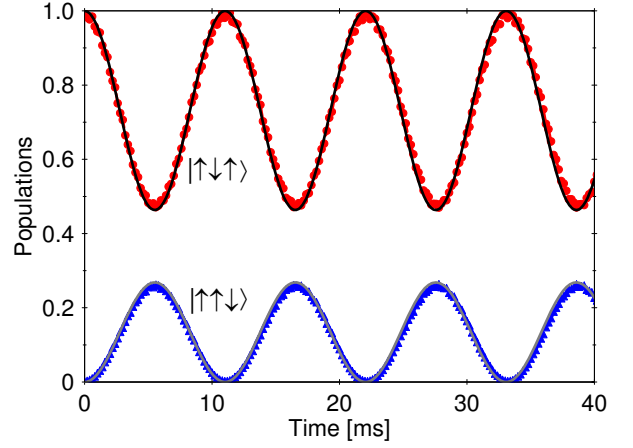}
\caption{(Color online) Superexchange interaction in a system of three ions couples the states $\left|\uparrow\downarrow\uparrow\right\rangle$, $|\left\uparrow\uparrow\downarrow\right\rangle$ and $\left|\downarrow\uparrow\uparrow\right\rangle$ according to the effective Hamiltonian (\ref{XXZ}). We compare the probability of finding the system in states $\left|\uparrow\downarrow\uparrow\right\rangle$ and $|\left\uparrow\uparrow\downarrow\right\rangle$ computed by the effective Hamiltonian (\ref{XXZ}) (red circles and blue triangles) and Hamiltonian (\ref{JCH}) (solid lines). The population of state $\left|\downarrow\uparrow\uparrow\right\rangle$ is indistinguishable from that of $\left|\uparrow\uparrow\downarrow\right\rangle$ and it is not shown in the figure. We assume axial trap frequency $\omega_{z}/2\pi=120$ kHz and aspect ratios $\omega_{y}/\omega_{x}=1.8$ and $\omega_{y}/\omega_{z}=100$. The parameters are set to $t_{1,2}^{x}/2\pi=0.86$ kHz, $t_{1,2}^{y}/2\pi=0.48$ kHz, $t_{1,3}^{x}/2\pi=0.1$ kHz, $t_{1,3}^{y}/2\pi=0.06$ kHz and $g_{x}/2\pi=19$ kHz, $g_{y}/2\pi=20$ kHz, $\delta/2\pi=-0.22$ kHz, which ensure that the system is in the strong-coupling regime.
The phonon detuning is set to $\delta\omega_{\beta}=\omega_{\beta,1}$, such that we have $\delta\omega_{x,1}-\delta\omega_{x,2}=2\pi\times8$ Hz and $\delta\omega_{y,1}-\delta\omega_{y,2}=2\pi\times4$ Hz.}
\label{fig3}
\end{figure}

A second-order hopping process to a state with two excitations in one site and none in another creates an effective spin-spin interaction between spin-$\frac{1}{2}$ systems on different lattice sites.
By calculating the matrix elements in Eq. (\ref{PertH}) we find that the resulting spin dynamics is described by the anisotropic XXZ Heisenberg Hamiltonian in the presence of external magnetic field,
\begin{equation}
\hat{H}_{\rm eff}=\sum_{j<k}K_{j,k}^{xy}(\sigma_{j}^{x}\sigma_{k}^{x}+\sigma_{j}^{y}\sigma_{k}^{y})
+\sum_{j<k}K_{j,k}^{z} \sigma_{j}^{z}\sigma_{k}^{z}+\sum_{j}H_{j}\sigma_{j}^{z},\label{XXZ}
\end{equation}
where $\sigma_{j}^{x}=(\left|\uparrow_{j}\right\rangle\left\langle\downarrow_{j}\right|+{\rm H.c})$, $\sigma_{j}^{y}=-i(\left|\uparrow_{j}\right\rangle\left\langle\downarrow_{j}\right|-{\rm H.c})$ and $\sigma_{j}^{z}=\left|\uparrow_{j}\right\rangle\left\langle\uparrow_{j}\right|-\left|\downarrow_{j}\right\rangle
\left\langle\downarrow_{j}\right|$ denote the corresponding spin operators of the system.
The couplings in Eq. (\ref{XXZ}) derived by second-order perturbation theory in the phonon hopping are given by
\bse\label{C}
\begin{align}
K_{j,k}^{xy} &=-t_{j,k}^{x}t_{j,k}^{y}\frac{2(\tan\zeta+\cot\zeta)+5}{8g_{y}(1+\tan\zeta)},\\
K_{j,k}^{z} &=\frac{(t_{j,k}^{x})^{2}(\tan\zeta-6\cot\zeta-4)+(t_{j,k}^{y})^{2}(\cot\zeta-6\tan\zeta-4)}
{16g_{y}(1+\tan\zeta)},\\
H_{j} &=-\frac{5}{8}\sum_{k\neq j}\Big[\frac{(t_{j,k}^{x})^{2}}{g_{x}}-\frac{(t_{j,k}^{y})^{2}}{g_{y}}\Big],
\end{align}
\ese
where we take $\delta=0$.
Off resonance, the expressions are too long to be presented here.
In Fig. \ref{fig3} we show the comparison between the JCHv Hamiltonian (\ref{JCH}) and the effective spin model (\ref{XXZ}) for a linear ion crystal with three ions.
The superexchange couplings cause oscillation between the initial state $\left|\uparrow\downarrow\uparrow\right\rangle$ and states $\left|\uparrow\uparrow\downarrow\right\rangle$, $\left|\downarrow\uparrow\uparrow\right\rangle$ according to the spin model (\ref{XXZ}). Obviously, the effective spin model matches the exact dynamics very accurately.

\begin{figure}[tb]
\includegraphics[width=0.45\textwidth]{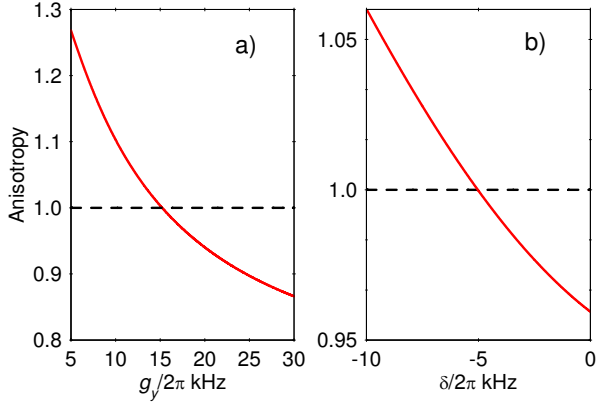}
\caption{(Color online). The anisotropy $\lambda_{1,2}=J_{1,2}^{z}/J_{1,2}^{xy}$ in a system of three ions. a) The anisotropy as a function of $g_{y}$. The parameter are set to $g_{x}/2\pi=12$ kHz, and $\delta/2\pi=-0.5$ kHz.
 b) We fixed the coupling $g_{y}/2\pi=18$ kHz and vary the detuning $\delta$. The hopping amplitudes are set to  $t_{1,2}^{x}/2\pi=0.5$ kHz, $t_{1,2}^{y}/2\pi=0.7$ kHz.}
\label{a}
\end{figure}

Finally, we note that the couplings (\ref{C}) can be tuned by adjusting the external parameters, namely, the axial trap frequency, the laser field intensities and the detuning.
For example, one could control the amount of spin-exchange anisotropy $\lambda_{i,j}=K_{i,j}^{z}/K_{i,j}^{xy}$ by varying the spin-phonon couplings or the detuning, as demonstrated in Fig. \ref{a}.
This allows us to choose the appropriate parameters such that an easy-axis $\lambda_{i,j}\geq 1$ or easy-plane $\lambda_{i,j}<1$ ferromagnets are realized \cite{Auerbach}.

\section{Heisenberg-like model with spin-$1$}\label{Spin1}

\begin{figure}[tb]
\includegraphics[width=0.45\textwidth]{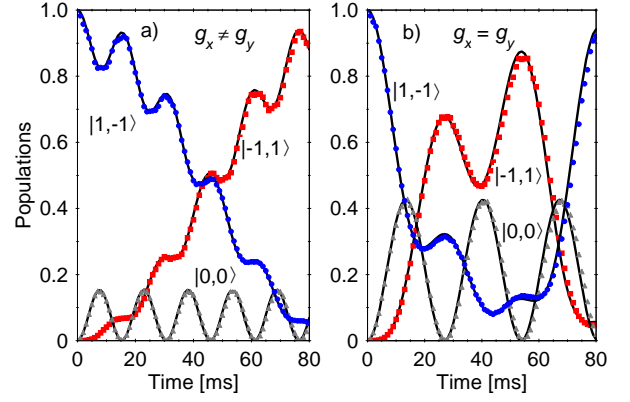}
\caption{(Color online). Coherent superexchange interaction in a system of two ions. a) We plot the time evolution of states $\left|1,-1\right\rangle$ (blue circles), $\left|-1,1\right\rangle$ (red squares) and $\left|0,0\right\rangle$ (grey triangles) according to the effective Heisenberg-like Hamiltonian (\ref{HS=1}) compared with the JCHv Hamiltonian (\ref{JCH}) (solid lines). The parameter are set to $g_{x}/2\pi=32$ kHz, $g_{y}/2\pi=34$ kHz, $\delta=0$, $t^{x}_{1,2}/2\pi=0.1$ kHz and $t^{y}_{1,2}/2\pi=0.17$ kHz. b) the same but the spin phonon couplings are set to $g_{x}/2\pi=34$ kHz and $g_{x}=g_{y}$.
}
\label{s1}
\end{figure}

A simple generalization of the interacting spin models with higher spins can be obtained by considering the case of two polaritonic excitations per site.
Then the low-lying energy manifold of the Hamiltonian $\hat{H}_{\rm JC}$ consists of three eigenstates (\ref{n3}), which in the following will represent an effective spin $s=1$ system.
The energies of these states are degenerate for $g_{x}=g_{y}$, while for unequal couplings $g_{x}\neq g_{y}$ the degeneracy is lifted and due to the non-linearity in the energy spectrum, the differences $E_{1}-E_{0}$ and $E_{0}-E_{-1}$ are not equidistant.
In the strong-coupling regime, the second-order hopping processes to the states with three excitations in one site and one excitation in another create couplings between the states (\ref{n3}) at different lattice sites, which allow us to map the original Hamiltonian (\ref{JCH}) to an effective spin $s=1$ model.
After calculating the matrix elements, we arrive to the following highly anisotropic effective Heisenberg-like Hamiltonian (see, Appendix \ref{Derivation})
\begin{align}
\hat{H}_{\rm eff} &=
\sum_{j} [D_{j}(\hat{S}_{j}^{z})^{2} +B_{j}\hat{S}_{j}^{z}] \notag\\
 &+\sum_{j<k} J_{j,k}^{xy}(\hat{S}_{j}^{x}\hat{S}_{k}^{x}+\hat{S}_{j}^{y}\hat{S}_{k}^{y}) + \sum_{j<k} J_{j,k}^{z}\hat{S}_{j}^{z}\hat{S}_{k}^{z} \notag\\
&+\sum_{j<k}W_{j,k}[\hat{S}_{j}^{z}(\hat{S}_{k}^{z})^{2}+(\hat{S}_{j}^{z})^{2}\hat{S}_{k}^{z}]
+\sum_{j<k}V_{j,k}(\hat{S}_{j}^{z}\hat{S}_{k}^{z})^{2}\notag\\
&+\sum_{j<k} [v_{j,k}^{(1)}(\hat{S}_{j}^{z}\hat{S}_{j}^{+}\hat{S}_{k}^{-}\hat{S}_{k}^{z}+{\rm H.c.})+v_{j,k}^{(-1)}(\hat{S}_{j}^{z}\hat{S}_{j}^{-}\hat{S}_{k}^{+}\hat{S}_{k}^{z}+{\rm H.c.})] .
\label{HS=1}
\end{align}
Here $\hat{S}_{j}^{x}=\frac{1}{2}(\hat{S}_{j}^{+}+\hat{S}_{j}^{-})$, $\hat{S}_{j}^{y}=\frac{i}{2}(\hat{S}_{j}^{-}-\hat{S}_{j}^{+})$, and $\hat{S}_{j}^{z}=-i[\hat{S}_{j}^{x},\hat{S}_{j}^{y}]$ are the spin $s=1$ operators at site $j$.
The Hamiltonian (\ref{HS=1}) represents the spin $s=1$ Heisenberg model with Ising-like and single-ion anisotropy terms \cite{Auerbach}.
Such an anisotropy of the spin-spin interactions occurs due to the non-equidistance in the energies of spin $s=1$ system, which reflects into the matrix elements in Eq. (\ref{PertH}).
Indeed, for equal spin-phonon couplings ($g_{x}=g_{y}$) the degeneracy of the states (\ref{n3}) equalizes the superexchange interaction between the states and the effective Hamiltonian corresponds to the anisotropic ferromagnetic Heisenberg model in the presence of the external magnetic field,
\begin{equation}
\hat{H}_{\rm eff}=\sum_{j<k}\{J_{j,k}^{xy}(\hat{S}_{j}^{x}\hat{S}_{k}^{x}+\hat{S}_{j}^{y}\hat{S}_{k}^{y})
+J_{j,k}^{z}\hat{S}_{j}^{z}\hat{S}_{k}^{z}\}+\sum_{j}B_{j}\hat{S}_{j}^{z}.\label{HS=1i}
\end{equation}
In Fig. \ref{s1} we check the validity of the perturbative approach by comparing the effective Hamiltonian (\ref{HS=1}) with the JCHv Hamiltonian (\ref{JCH}) in a system of two ions.
For unequal couplings, Fig. \ref{s1}(a), the probabilities of finding the system in the states $\left|1,-1\right\rangle$, $\left|0,0\right\rangle$, and $\left|-1,1\right\rangle$ evolve according to Eq. (\ref{HS=1}), while in the case of equal couplings the spin evolution is governed by the effective Hamiltonian (\ref{HS=1i}), Fig. \ref{s1}(b).

The Heisenberg-like model (\ref{HS=1}) presented here can be considered as a generalization of the highly anisotropic spin-1 models recently investigated in a system of ultracold dipolar molecules loaded in a one-dimensional optical lattice \cite{Kestner_2011,Dalmonte_2011}.
However, to the best of our knowledge, there is no extensive study of the entire phase diagram of our highly anisotropic spin-1 model.
As was pointed out in Refs.~\cite{Gu_2009,Pollmann_2010,Pollmann_2012}, the gapped phases of any one-dimensional spin model can be classified by its symmetry group.
An example of such topological phases is the Haldane phase, which appears in one-dimensional integer-spin chains \cite{Haldane}.
The latter are characterized with nonzero excitation gaps and exponentially decaying spin correlation functions.
The stability of the Haldane phase crucially depends on the protection of an appropriate set of symmetries.
Since our model contains an odd number of spin operators, the only discrete symmetry of the Hamiltonian (\ref{HS=1}) is the rotation by $\pi$ around the $z$ axis, which takes $\hat{S}_{j}^{x,y}\rightarrow -\hat{S}_{j}^{x,y}$ and $\hat{S}_{j}^{z}\rightarrow \hat{S}_{j}^{z}$, while the Hamiltonian (\ref{HS=1i}) obeys an additional symmetry, which is a rotation by $\pi$ around the $y$ axis and time-reversal such that $\hat{S}_{j}^{x,z}\rightarrow \hat{S}_{j}^{x,z}$ and $\hat{S}_{j}^{y}\rightarrow -\hat{S}_{j}^{y}$.
As was pointed out in Ref.~\cite{Kestner_2011}, such spin-$1$ models may exhibit novel nontrivial topological order.

Although the spin couplings $J_{j,k}^{xy}<0$ and $J_{j,k}^{z}<0$ support ferromagnetic ground-state order we may use the duality between ferro- and antiferromagnetic models, i.e. $\hat{H}_{\rm AF}=-\hat{H}_{\rm F}$ \cite{Ripoll}.
The latter implies that the highest energy state of the ferromagnetic model is in fact the ground state of the corresponding antiferromagnetic Hamiltonian.
The key observation here is that one could switch between both spin models (\ref{HS=1}) and (\ref{HS=1i}) by controlling the intensities of the laser beams.
For example, the preparation can start by setting $g_{x}=g_{y}$ and $t_{j,k}^{x}\gg t_{j,k}^{y}$, which implies $|J_{j,k}^{z}|\gg |J_{j,k}^{xy}|$, and prepare the antiparallel spin configuration.
Such a state can be realized by ground-state cooling of the radial vibrational modes and pumping the internal ion states to $|g\rangle_{j}$.
The antiparallel configuration between states $\left|1\right\rangle$ and $\left|-1\right\rangle$ can be created by noting that for large negative detuning ($|\delta|\gg g_{\beta}$) the polaritonic nature of the states is reduced to $\left|1\right\rangle\approx|e_{1},1_{x},0_{y}\rangle$ and $\left|-1\right\rangle\approx|e_{2},0_{x},1_{y}\rangle$, see Eq. (\ref{n3}).
The latter states can be created by $\pi$ pulses that are resonant with the respective blue-sideband transitions \cite{Toyoda2013}.
Once the initial state is prepared one could lower $\delta$ and induce unequal spin-phonon couplings.
The superexchange interaction can be probed by letting the system to evolve and then measure either the local phonon number or the internal ion population.

\section{Conclusions}\label{conclusion}
We have shown that a laser-driven linear ion crystal can realize the JCHv model.
We have studied the strongly coupled regime where the JCHv model can be mapped to effective spin models.
We have considered the case of one and two polaritonic excitations per site, which represent our effective spin-$\frac{1}{2}$ and spin-$1$ systems.
The underlying mechanism that creates the spin-spin couplings is the Heisenberg superexchange interaction, which can be controlled by the trap frequencies, the laser intensity and the detuning.


\acknowledgments

This work has been supported by the EC Seventh Framework Programme under Grant Agreement No. 270843 (iQIT) and by Sofia University Grant No. 084/2014.

\appendix

\section{Implementation of JCHv model in an optical V-type system}\label{Impl_LI}

Here we consider an alternative implementation of the JCHv model with ions, which possess three long-lived internal states in the microwave domain.
In order to induce the JC couplings between the ground state and the two excited states we assume that the ion crystal interacts with a time-varying magnetic field quadrupole,
\begin{equation}
\vec{B}(t)=bf(t)(x\vec{e}_{x}-y\vec{e}_{y}).
\end{equation}
We note that such a magnetic field quadrupole was used recently for implementation of entangling operations in an ion trap \cite{Ospelkaus_nature2011}. The total Hamiltonian is
\begin{equation}
\hat{H}=\hat{H}_{xy}+\hat{H}_{s}+\hat{H}_{\rm I},\label{H}
\end{equation}
where
\be
\hat{H}_{s}=\sum_{j=1}^{N}[\omega_{e,1}|e_{1,j}\rangle \langle e_{1,j}|+\omega_{e,2}|e_{2,j}\rangle \langle e_{2,j}|].
 \ee
 The interaction between the ionic internal states and the magnetic field is described by $\hat{H}_{\rm I}=-\sum_{j}(\vec{\mu}_{j}^{ge_{1}}+\vec{\mu}_{j}^{ge_{2}})\vec{B}_{j}$. Here $\vec{\mu}^{ge_{a}}$ ($a=1,2$) is the magnetic dipole moment operator between states $|g\rangle\leftrightarrow |e_{a}\rangle$, respectively.
The latter can be expressed as $\vec{\mu}^{ge_{a}}=\mu_{x}^{ge_{a}}(|e_{a}\rangle\langle g|+{\rm H.c})-i\mu_{y}^{ge_{a}}(|e_{a}\rangle\langle g|-{\rm H.c})$ and in the following we assume the condition
 $\mu_{x}^{ge_{a}}=\mu_{y}^{ge_{a}}=\mu^{ge_{a}}$.
 Using this, we write the interaction Hamiltonian as
\begin{eqnarray}
\hat{H}_{\rm I}&=&-b f(t)\sum_{a=1,2}\sum_{j=1}^{N}\mu^{ge_{a}}[\delta \hat{r}_{x,j}(|e_{a,j}\rangle\langle g_{j}|+{\rm H.c.})\notag\\
&&-i\delta \hat{r}_{y,j}(|e_{a,j}\rangle\langle g_{j}|-{\rm H.c.})].
\end{eqnarray}
Clearly, the magnetic field mediates a coupling between the internal levels and motional displacements.
In order to induce the desired JC type interaction we use a periodic driving field
\be
f(t)=\cos\nu_{1}t+\cos\nu_{2}t
\ee
 with frequencies
 \bse
 \begin{align}
 \nu_{1} &=\omega_{e,1}-\omega_{0}-(\bar{\omega}_{x}-\Delta),\\
 \nu_{2} &=\omega_{e,2}-\omega_{0}-(\bar{\omega}_{y}-\Delta),
 \end{align}\ese
  with $\bar{\omega}_{\beta}=\omega_{\beta}+\delta\omega_{\beta}$.
Such a choice of driving frequencies allows to excite a JC interaction between $|g\rangle\rightarrow |e_{1(2)}\rangle$ states with creation of $x(y)$ phonons, respectively.
Indeed, by performing a unitary transformation of Hamiltonian (\ref{H}) into a rotating frame with respect to
\begin{align}
\hat{U}(t) =& \exp[-i\sum_{j}\{\sum_{a=1}^{2}(\omega_{e,a}-\omega_{0})|e_{a,j}\rangle \langle e_{a,j}| \notag\\
&+\sum_{\beta}(\bar{\omega}_{\beta}-\Delta)\hat{a}_{\beta,j}^{\dag}\hat{a}_{\beta,j}\}t],
 \end{align}
 we find $\hat{H}_{\rm JCHv}=\hat{U}^{\dag}\hat{H}\hat{U}-i\hat{U}^{\dag}\partial_{t}\hat{U}$, or
\bse
\begin{align}
\hat{H}_{\rm JCHv}=& \hat{H}_{\rm JC}+\hat{H}_{\rm b}, \\
\hat{H}_{\rm JC}=& \sum_{j}\sum_{\beta}\Delta_{\beta,j}\hat{a}_{\beta,j}^{\dag}\hat{a}_{\beta,j}
+\sum_{j} \omega_{0}(|e_{1,j}\rangle\langle e_{1,j}|+|e_{2,j}\rangle\langle e_{2,j}|)\notag\\
&+\sum_{j} g_{x}(\hat{a}_{x,j}|e_{1,j}\rangle \langle g_{j}|+{\rm H.c}) \notag\\
&+\sum_{j} g_{y}(\hat{a}_{y,j}|e_{2,j}\rangle \langle g_{j}|+{\rm H.c}) ,\label{JCHm}
\end{align}
\ese
where $\Delta_{\beta,j}=\Delta+\delta\omega_{\beta,j}-\delta\omega_{\beta}$.
The couplings between the internal states and the local phonon states are $g_{x}=-b\mu^{ge_{1}}/\sqrt{2m\omega_{x}}$ and $g_{y}=-b\mu^{ge_{2}}/\sqrt{2m\omega_{y}}$, respectively.
Finally, in Hamiltonian (\ref{JCHm}) fast-rotating terms are neglected, which is justified as long as $g_{\beta}\ll|\omega_{x}\pm\omega_{y}|$, $\omega_{e,a}$.

\section{Derivation of spin-1 Hamiltonian}\label{Derivation}

By using Eq. (\ref{PertH}) we find that the spin-spin interaction gives rise to the effective Hamiltonian
\begin{equation}
\hat{H}_{j,k}=(\hat{H}_{\rm JC})_{j,k}\delta_{j,k}+\hat{H}_{j,k}^{xy}+\hat{H}_{j,k}^{z}\label{Eff},
\end{equation}
with $\hat{H}_{\rm eff}=\sum_{j<k}\hat{H}_{j,k}$.
The first term can be written as
\begin{equation}
\hat{H}_{\rm JC}=\sum_{j} (E_{1}\hat{X}_{j}^{11}+E_{0}\hat{X}_{j}^{00}+E_{-1}\hat{X}_{j}^{-1-1}),
\end{equation}
where we truncate the Hilbert space only to the three lowest eigenstates of $\hat{H}_{\rm JC}$.
Here we have defined the Hubbard operators $\hat{X}_{j}^{ab}=|a_{j}\rangle\langle b_{j}|$ ($a,b=1,0,-1$).
Expressing the Hubbard operators in terms of spin $s=1$ operators, $\hat{X}_{j}^{11}=\frac{1}{2}[(\hat{S}_{j}^{z})^{2}+\hat{S}_{j}^{z}]$, $\hat{X}_{j}^{00}=\hat{\mathbb{I}}_{j}-(\hat{S}_{j}^{z})^{2}$ and $\hat{X}_{j}^{-1-1}=\frac{1}{2}[(\hat{S}_{j}^{z})^{2}-\hat{S}_{j}^{z}]$, we obtain
\begin{equation}
\hat{H}_{\rm JC}=\frac{1}{2}\sum_{j} [\hat{S}_{j}^{z}(E_{1}-E_{-1})+(\hat{S}_{j}^{z})^{2}(E_{1}+E_{-1}-2E_{0})+2E_{0}\hat{\mathbb{I}}_{j}].\label{JC}
\end{equation}
Note that in the isotropic case $g_{x}=g_{y}$ the energies are degenerate; hence the term (\ref{JC}) yields only a global phase.
The last two terms in Eq. (\ref{Eff}) arise due to the second-order hopping processes and can be written as
\bse\label{H_S}
\begin{align}
\hat{H}_{j,k}^{xy} =&T_{j,k}^{(1)}(\hat{X}_{j}^{10}\hat{X}_{k}^{01}+{\rm H.c.})+
T_{j,k}^{(-1)}(\hat{X}_{j}^{-10}\hat{X}_{k}^{0-1}+{\rm H.c.})\notag\\
&+T_{j,k}^{(0)}(\hat{X}_{j}^{10}\hat{X}_{k}^{-10}+\hat{X}_{j}^{-10}\hat{X}_{k}^{10}+{\rm H.c.})\label{H_transition}
\\
\hat{H}_{j,k}^{z} =&T_{j,k}^{(1,1)}\hat{X}_{j}^{11}\hat{X}_{k}^{11}+T_{j,k}^{(0,0)}\hat{X}_{j}^{00}\hat{X}_{k}^{00}
+T_{j,k}^{(-1,-1)}\hat{X}_{j}^{-1-1}\hat{X}_{k}^{-1-1}\notag\\
&+T_{j,k}^{(1,0)}(\hat{X}_{j}^{11}\hat{X}_{k}^{00}+\hat{X}_{j}^{00}\hat{X}_{k}^{11}) \notag\\
&+T_{j,k}^{(0,-1)}(\hat{X}_{j}^{00}\hat{X}_{k}^{-1-1}+\hat{X}_{j}^{-1-1}\hat{X}_{k}^{00})\notag\\
&+T_{j,k}^{(1,-1)}(\hat{X}_{j}^{11}\hat{X}_{k}^{-1-1}+\hat{X}_{j}^{-1-1}\hat{X}_{k}^{11}).
\end{align}
\ese
The coupling coefficients in (\ref{H_S}) are the respective matrix elements of the second term in Eq. (\ref{PertH}). The Hamiltonian $\hat{H}_{j,k}^{xy}$ describes the transition probabilities between different spin states. In terms of spin $s=1$ operators it is expressed as
\begin{align}
\hat{H}_{j,k}^{xy} =& v_{j,k}^{(1)}(\hat{S}_{j}^{z}\hat{S}_{j}^{+}\hat{S}_{k}^{-}\hat{S}_{k}^{z}+{\rm H.c.})+v_{j,k}^{(-1)}(\hat{S}_{j}^{z}\hat{S}_{j}^{-}\hat{S}_{k}^{+}\hat{S}_{k}^{z}+{\rm H.c.})\notag\\
& +J_{j,k}^{xy}(\hat{S}_{j}^{x}\hat{S}_{k}^{x}+\hat{S}_{j}^{y}\hat{S}_{k}^{y}),\label{H_transitionS}
\end{align}
with the couplings given by
$v_{j,k}^{(1)} =\tfrac{1}{2}(T_{j,k}^{(1)}-T_{j,k}^{(0)})$,
$v_{j,k}^{(-1)} =\tfrac{1}{2}(T_{j,k}^{(-1)}-T_{j,k}^{(0)})$,
$J_{j,k}^{xy} =T_{j,k}^{(0)}$.
The spin-dependent energy corrections due to the hopping events are described by $\hat{H}_{j,k}^{z}$, which can be written as
\begin{align}
\hat{H}_{j,k}^{z} =&
B_{j,k}(\hat{S}_{j}^{z}+\hat{S}_{k}^{z})
+D_{j,k}[(\hat{S}_{j}^{z})^{2}+(\hat{S}_{k}^{z})^{2}]
+J_{j,k}^{z}\hat{S}_{j}^{z}\hat{S}_{k}^{z}
\notag\\
&+W_{j,k}[\hat{S}_{j}^{z}(\hat{S}_{k}^{z})^{2}+(\hat{S}_{j}^{z})^{2}\hat{S}_{k}^{z}]
+V_{j,k}(\hat{S}_{j}^{z}\hat{S}_{k}^{z})^{2}
\label{H_energyS}.
\end{align}
The coupling coefficients are given by
\bse
\begin{align}
J_{j,k}^{z} =& \tfrac{1}{4} \big[T_{j,k}^{(1,1)}+T_{j,k}^{(-1,-1)}-2T_{j,k}^{(1,-1)}\big],\\
W_{j,k} =& \tfrac{1}{4} \big[T_{j,k}^{(1,1)} -T_{j,k}^{(-1,-1)}+2T_{j,k}^{(0,-1)}-2T_{j,k}^{(1,0)}\big],\\
V_{j,k} =& \tfrac{1}{4} \big[T_{j,k}^{(1,1)}+T_{j,k}^{(-1,-1)}+2T_{j,k}^{(1,-1)}+4T_{j,k}^{(0,0)}\notag\\
&-4T_{j,k}^{(1,0)}-4T_{j,k}^{(0,-1)}\big],\\
D_{j,k} =& \tfrac{1}{2} \big[T_{j,k}^{(1,0)}+T_{j,k}^{(0,-1)}-2T_{j,k}^{(0,0)}\big],\\
B_{j,k} =& \tfrac{1}{2} \big[T_{j,k}^{(1,0)}-T_{j,k}^{(0,-1)}\big].
\end{align}
\ese
Obviously, the single-ion anisotropy term and the external magnetic-field term in Eq. (\ref{HS=1}) are given by
\bse
\begin{align}
D_{j} &=\tfrac{1}{2}(E_{1}+E_{-1}-2E_{0})+\sum_{k\neq j}D_{j,k},\\
B_{j} &=\tfrac{1}{2}(E_{1}-E_{-1})+\sum_{k\neq j}B_{j,k}.
\end{align}
\ese
For $g_{x}=g_{y}$ the spin coefficients in Eq. (\ref{H_transition}) become equal, $T_{j,k}^{(1)}=T_{j,k}^{(-1)}=T_{j,k}^{(0)}$, and the Hamiltonian (\ref{H_transitionS}) reduces to
\begin{equation}
\hat{H}_{j,k}^{xy}=J_{j,k}^{xy}(\hat{S}_{j}^{x}\hat{S}_{k}^{x}+\hat{S}_{j}^{y}\hat{S}_{k}^{y}),
\end{equation}
with
\be
J_{j,k}^{xy}=-\frac{123\sqrt{2}}{7g_{x}}t_{j,k}^{x}t_{j,k}^{y},
\ee
 where we take $\delta=0$.
The same simplification holds for Eq. (\ref{H_energyS}),
\begin{equation}
\hat{H}_{j,k}^{z}=J_{j,k}\hat{S}_{j}^{z}\hat{S}_{k}^{z}+B_{j,k}(\hat{S}_{j}^{z}+\hat{S}_{k}^{z}),
\end{equation}
with
\begin{align}
J_{j,k}^{z} &=-\frac{123}{7\sqrt{2}g_{x}} [(t_{j,k}^{x})^{2}+(t_{j,k}^{y})^{2}],\\
B_{j,k} &=-\frac{53}{2\sqrt{2}g_{x}} [(t_{j,k}^{x})^{2}-(t_{j,k}^{y})^{2}].
\end{align}


\end{document}